\documentclass[12pt]{article}
\let\section=\subsection     \let\subsection=\subsubsection        
\usepackage{graphicx}

\begin{document}
\begin{center}
{\large \bf  Statistical coalescence model}\\[2mm]
{\large \bf of $J/\psi$ production at the SPS and RHIC}\\[5mm] 
  Mark I. Gorenstein \\[5mm]
   {\small \it   Bogolyubov Institute for Theoretical Physics,
Kiev, Ukraine\\
Institut f\"ur Theoretische Physik, Universit\"at  Frankfurt,
Germany
 \\[8mm] }
\end{center}

\begin{abstract}\noindent
A recently developed statistical coalescence model of the
$J/\psi$ 
production is presented.
The NA50 data on the $J/\psi$ suppression pattern and the transverse mass
spectra in Pb+Pb collisions at the SPS are analyzed. The model predictions
for the RHIC energies are formulated. The measurements of $J/\psi$ in
Au+Au collisions at RHIC
are crucial for disentangling the different scenarios of
charmonia formation.
\end{abstract}

\section{Introduction}
The experimental studies of
nucleus-nucleus
(A+A) collisions at the
SPS and RHIC  
provide a rich information on hadron observables (multiplicities and
momentum spectra). An  extensive experimental program is motivated by a
possibility to create a new state of matter -- the quark gluon plasma
(QGP) -- in a laboratory.
Any theoretical description of the QGP requires a macroscopic approach:
statistical mechanics to describe the static properties of the matter
and hydrodynamics to describe the system evolution.
The equilibrium  hadron gas (HG) model describes remarkably well
the hadron multiplicities measured
in Pb+Pb and Au+Au collisions at top SPS ($\sqrt{s}=17$~GeV per
nucleon pair) \cite{HG} and
RHIC ($\sqrt{s}=130$~GeV) \cite{HG1} energies,
where the creation of QGP is expected.

The HG model assumes the
following formula for the hadron thermal multiplicities
in the grand canonical ensemble (g.c.e.):
\begin{equation}\label{stat}
N_j~=~\frac{d_j~V}{2\pi^2} ~
\int_0^{\infty}p^2dp~\left[\exp\left(
\frac{\sqrt{p^2+m_{j}^2} - \mu_j}{T}\right)~\pm~1\right]^{-1}~,
\end{equation}
where $V$ and $T$ are the system volume and temperature, respectively,
$m_j$ and $d_j$ denote particle masses and
degeneracy
factors.
The particle chemical
potential $\mu_j$ in Eq.(\ref{stat})
is defined as
%\begin{equation}\label{mui}
$\mu_j=b_j\mu_B+s_j\mu_S+c_j\mu_C$~,
%\end{equation}
where $b_j,s_j,c_j$ denote the baryonic number strangeness and
charm of particle $j$. The baryonic chemical potential $\mu_B$
regulates the baryonic density of the HG system whereas
strange $\mu_S$ and charm $\mu_C$ chemical potentials should be found
from the requirement of zero value for the total strangeness and charm   
in the system (in our consideration we neglect small effects
of a non-zero electrical chemical potential).
The total multiplicities $N_j^{tot}$ in the HG model
include the resonance decay
contributions:
\begin{equation}\label{dec}
N_j^{tot}~=~N_j ~+~\sum_R Br(R \rightarrow j) N_R~,
\end{equation}
where $Br(R \rightarrow j)$ are the corresponding decay
branching ratios.
The hadron yield ratios $N_j^{tot}/N_i^{tot}$
in the g.c.e. are the functions of $T$ and $\mu_B$ and are
independent of the volume parameter $V$.

The temperature parameters extracted from fitting the multiplicity data
are approximately the same for the SPS and RHIC energies:
$T=170\pm 10$~MeV \cite{HG,HG1}.
This value is close to an
estimate of the temperature
$T_c$ for the QGP--HG  transition
obtained in Lattice QCD simulations at zero baryonic
density (see e.g. \cite{Karsch}).
One may therefore argue that QGP formed at the early stage of
A+A reaction hadronizes into
a locally equilibrated HG,
and the  chemical composition of this hadron gas is weakly affected by the
subsequent hadron rescatterings.
%Therefore, at the SPS and RHIC the so-called ``chemical freeze-out''
%is rather close to (or even coincide with) the hadronization transition.

\section{ Statistical coalescence model of  $J/\psi$
production}
The production of charmonium states $J/\psi$ and $\psi^{\prime}$ have been
measured in A+A
collisions at CERN SPS over the last 15 years by the NA38 and NA50
Collaborations.
These studies were mainly motivated by the theoretical suggestion
\cite{Satz1} to use
$J/\psi$ meson as a probe for deconfinement in  A+A
collisions. 

Recently  the thermal  model \cite{Ga1}
and the statistical coalescence model (SCM) \cite{Br1,Go}
(see also \cite{Go1,Go2,Go3})
for the charmonium production in
A+A collisions were formulated.
The total $J/\psi$ multiplicity, $N_{J/\psi}^{tot}$, in
the thermal model \cite{Ga1}
is given by Eq.~(\ref{dec}), 
where $N_{J/\psi}$, $N_{\psi^{\prime}}$, $N_{\chi_1}$, $N_{\chi_2}$
are calculated according to Eq.(\ref{stat}) and $Br(\psi^{\prime})\cong
0.54$, $Br(\chi_1)\cong 0.27$, $Br(\chi_2)\cong 0.14$ are the decay
branching ratios of the excited charmonium states into $J/\psi$.
The thermal model \cite{Ga1} predicts that at high collision energies the
$J/\psi$ to $\pi$ ratio is independent of $\sqrt{s}$ and the number of
nucleon participants $N_p$.
This is because both
$\langle J/\psi \rangle$ and $\langle \pi \rangle$
multiplicities
are proportional to the system volume and 
the hadronization temperature 
is expected to be approximately constant
at high collision energies.

The SCM
\cite{Br1,Go} assumes that charmonium states
are formed at the hadronization stage.
This is similar to the thermal model \cite{Ga1}.
However, in the SCM the  charmonium states are
produced via a coalescence of
$c$ and $\overline{c}$ (anti)quarks created
by the hard parton collisions at the early
stage of A+A reaction. 
The number of created $c\overline{c}$
pairs, $N_{ c\overline{c}}$, in hard parton collisions, differs
in general
from the result
expected in the  equilibrium HG.
One needs then a {\it charm enhancement} factor
 $\gamma_c$  \cite{Br1}\footnote{
This is formally analogous to the
introduction of the {\it strangeness suppression} factor $\gamma_s < 1$
\cite{Raf1}
in the HG model.}
to
adjust the thermal HG results to the required 
average number, $\langle N_{c\overline{c}} \rangle $, of $c\overline{c}$
pairs. 
The open charm hadron yield
is enhanced
by a factor $\gamma_c$ and charmonium
yield by a factor $\gamma_c^2$ in comparison
with the equilibrium HG predictions.
This leads to a difference between
the thermal model and SCM predictions for charmonia multiplicities. 

The canonical ensemble (c.e.) formulation of the SCM
is \cite{Go}:
\begin{equation}\label{Ncc1}
\langle N_{c\overline{c}}\rangle~=~\frac{1}{2}~
\gamma_c~N_O~\frac{I_1(\gamma_c N_O)}{I_0(\gamma_cN_O)}~
+~\gamma_c^2~N_{H}~,
\end{equation}
where $N_{H}$ and $N_O$
are the HG
multiplicities of all particles with hidden and open charm, respectively.
The canonical suppression factor $I_1/I_0$ in Eq.(\ref{Ncc1})
is due to the exact charm conservation (see e.g., \cite{ce},
$I_0$ and $I_1$ are the modified Bessel functions).

If $\langle N_{c\overline{c}}\rangle$ is known, Eq.(\ref{Ncc1}) can be 
used
to find the charm enhancement
factor $\gamma_c$ and calculate then the $J/\psi$ multiplicity:
\begin{equation}\label{Npsi}
\langle J/\psi \rangle~= ~\gamma_c^2~N_{J/\psi}^{tot}~,
\end{equation}
where $N_{J/\psi}^{tot}$ is the total (thermal plus excited charmonium
decays) HG $J/\psi$ multiplicity.

Note that the second term in the right-hand side of Eq.(\ref{Ncc1}) gives
only a tiny
correction to the first term, i.e.  most  of the created
$c\overline{c}$ pairs are transformed into the open charm hadrons.
 Eqs.(\ref{Ncc1},\ref{Npsi}) lead then to:
\begin{equation}\label{Npsi1} 
\langle J/\psi \rangle~\cong~ 
\langle N_{c\overline{c}}\rangle~~ 
\frac{ N_{J/\psi}^{tot}}{(N_O/2)^2}~, ~~~
\langle N_{c\overline{c}}\rangle << 1~;
\end{equation}
\begin{equation}\label{Npsi2}
\langle J/\psi \rangle~\cong~        
\langle N_{c\overline{c}}\rangle^2~  
\frac{ N_{J/\psi}^{tot}}{(N_O/2)^2}~, ~~~ 
\langle N_{c\overline{c}}\rangle >> 1~. 
\end{equation}

Eq.(\ref{Ncc1}) assumes an exact conservation of
$N_c-N_{\bar{c}}\equiv 0$
and `statistical fluctuations' of
$N_{c\overline{c}}\equiv (N_c+N_{\bar{c}})/2$ numbers.
A more accurate treatment with `dynamical' (Poisson-like) distribution
of 
$N_{c\overline{c}}$ leads to \cite{Go2}:
\begin{equation}\label{Jpsi}
\langle J/\psi \rangle ~ \cong~
\langle N_{c\bar{c}} \rangle~
\left( 1 + \langle N_{c\bar{c}}\rangle \right)~
\frac{N_{J/\psi}^{tot}}{(N_O/2)^2} ~.
\end{equation}
Eq.(\ref{Jpsi}) coincides with Eqs.(\ref{Ncc1},\ref{Npsi}) at both  
$\langle N_{c\overline{c}}\rangle << 1$ and $\langle
N_{c\overline{c}}\rangle >> 1$ limits (see
Eqs.(\ref{Npsi1},\ref{Npsi2})). This is because at  
$\langle N_{c\overline{c}}\rangle << 1$ the probabilities to create
zero, $P(0)\cong 1- \langle N_{c\overline{c}}\rangle$, and one,
$P(1)\cong \langle N_{c\overline{c}}\rangle$, $c\overline{c}$ pairs 
are only
important. On the other hand, at $\langle N_{c\overline{c}}\rangle >> 1$
the statistical and dynamical
fluctuations of $N_{c\overline{c}}$ both obey the Poisson
law distribution.   

\section{ $J/\psi$ suppression in Pb+Pb collisions at the
SPS}
The centrality dependence of $\langle N_{c\bar{c}} \rangle$ 
in A+B nucleus-nucleus collisions can
be calculated in Glauber's approach,
$\langle N_{c\bar{c}} \rangle (b) = \sigma^{NN}_{c\bar{c}} T_{AB}(b)$,
where $b$ is the impact parameter,
$T_{AB}(b)$ is the nuclear overlap function and
$\sigma^{NN}_{c\bar{c}}$ is the $c\bar{c}$ production cross section
for nucleon-nucleon collisions. As discussed in Ref.\cite{Go1},
the deconfined medium can substantially modify charm production
at the SPS, i.e.,
$\sigma^{NN}_{c\bar{c}}$ in A+B collisions can be different from
the corresponding cross section measured in a nucleon-nucleon collision
experiment. The present analysis treats $\sigma^{NN}_{c\bar{c}}$
at the SPS energy as
a free parameter. Its value is fixed by fitting the NA50 data.

In the NA50 experiments \cite{NA50} the Drell-Yan muon pair multiplicity
(either
measured
or calculated from the minimum bias data) is used as a reference for
the $J/\psi$ ``suppression pattern''. 
The number of Drell-Yan pairs is also proportional to the number of
primary
nucleon-nucleon collisions:
$\langle DY' \rangle (b) = \sigma^{NN}_{DY'} T_{AB}(b)$,
where $\sigma^{NN}_{DY'}$ is the nucleon-nucleon
production cross section of $\mu^+ \mu^-$
Drell-Yan pairs. The prime means that the pairs should satisfy
the kinematical conditions of the NA50 spectrometer.
As the Drell-Yan cross section is isospin dependent, an average value
is used:
$\sigma^{NN}_{DY'}=\sigma^{AB}_{DY'}/(AB)$.
For the case of Pb+Pb collisions,
$A=B=208$ and
$\sigma^{PbPb}_{DY'} = 1.49 \pm 0.13 $ $\mu$b \cite{NA50}.

The quantity to be studied is the ratio
\begin{eqnarray}\label{Rb}
& & R_{DY}(b) ~ \equiv ~ \frac{\eta B^{J/\psi}_{\mu\mu} \langle J/\psi
\rangle (b)}
{\langle DY' \rangle (b)} \\
& =& ~
\eta ~B^{J/\psi}_{\mu\mu}~
\frac{\sigma^{NN}_{c\bar{c}}}{\sigma^{NN}_{DY'}}~~
\frac{\left( 1 +  \sigma^{NN}_{c\bar{c}} T_{AB}(b) \right)}{N_p(b)}~~
\frac{n_{J/\psi}^{tot}(T,\mu_B) n_B(T,\mu_B)}{(n_O(T,\mu_B)/2)^2}~.
 \nonumber 
\end{eqnarray}
$B^{J/\psi}_{\mu\mu}\cong 0.0588$ is the
decay probability of $J/\psi$ into $\mu^+\mu^-$.
Only the fraction $\eta$ of $\mu^+\mu^-$ pairs
satisfying the kinematical conditions
of the NA50 spectrometer
can be registered.
We treat $\eta$ in Eq.(\ref{Rb}) as one more model free parameter.

In the NA50 experiment, the neutral transversal energy $E_T$ of produced
particles
was used to measure centrality of the collisions. This variable,
however, provides a reliable measure of the centrality only if it does
not exceed a certain maximum value: $E_T < 100$ GeV
(see  Ref.\cite{Go2}). 
The average number of  participants is a linear function of the 
transversal energy, $\overline{N}_p =E_T/q$, 
in the domain $E_T < 100$~GeV ($q\cong0.274$~GeV). 
At $E_T > 100$~GeV the number of
$\overline{N}_p$
does not change essentially as $E_T$ grows. Therefore, the data at $E_T >
100$ GeV do not represent a centrality dependence of the $J/\psi$
suppression pattern 
but rather its dependence on fluctuations of the
stopping energy at approximately {\it fixed} number of participants.
The influence of such fluctuations on $J/\psi$ multiplicity can be
studied
in the framework of the SCM model \cite{Ko}.
\begin{center}
   \includegraphics[width=8cm,height=7cm,angle=0]{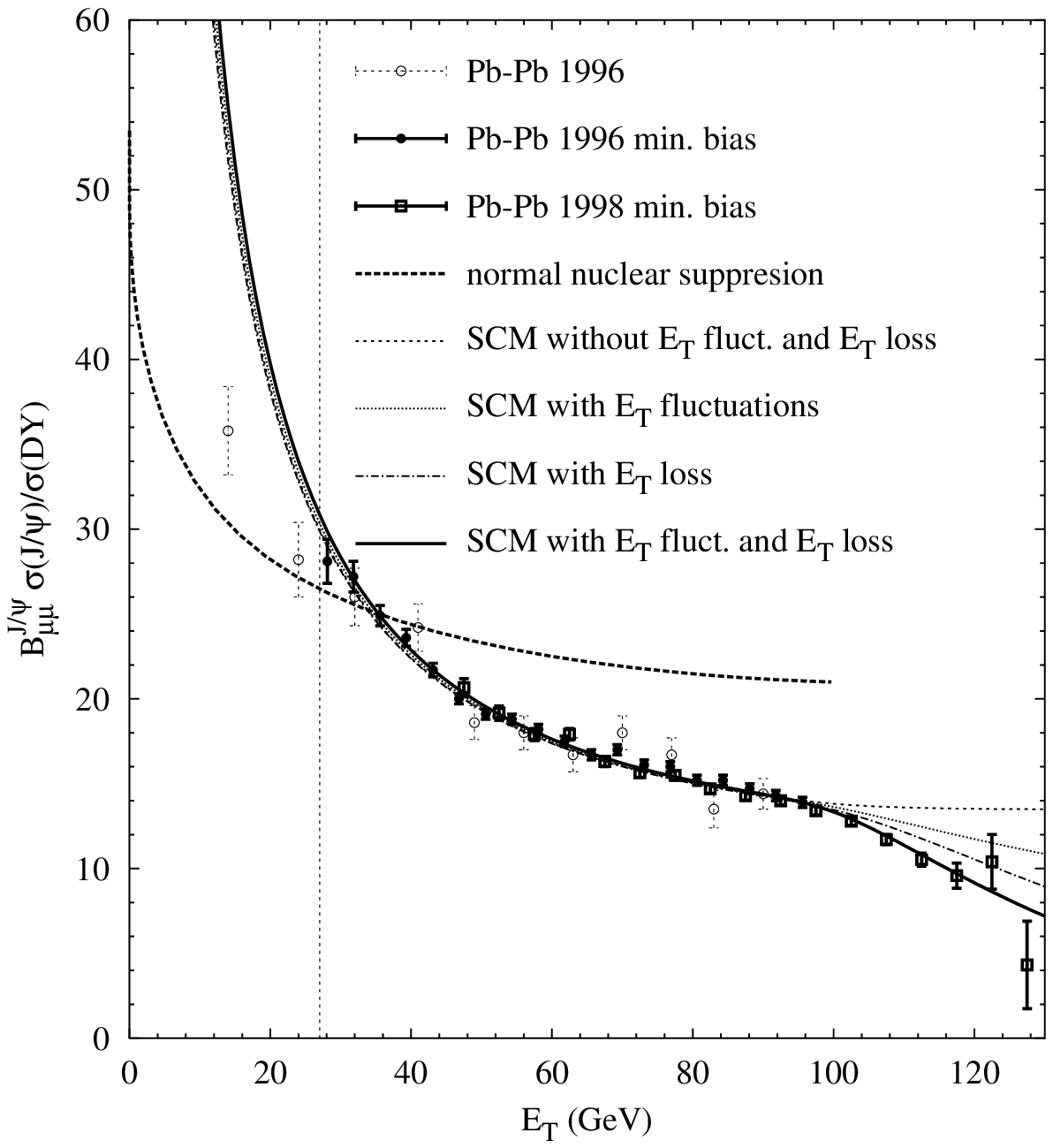}\\
   \parbox{14cm}
%        {\centerline
{\footnotesize
        Fig.~1: 
 The dependence of the $J/\psi$ to Drell-Yan ratio on the
transversal energy. The normal nuclear suppression curve is obtained at
$\sigma_{abs}=6.4$~mb. The SCM lines are calculated using
Eq.(\ref{Rb}). The
vertical line shows the applicability domain of
the SCM, $N_p>100$. 
}
\end{center}
The SCM fit of the NA50 data with Eq.(\ref{Rb}) is shown in Fig.~1
(see Refs.~\cite{Go2,Ko} for details). 
The model parameters are:
\begin{equation}\label{param}
\sigma^{NN}_{c\bar{c}}~\cong~35.7~\mu\mbox{b} ~,~~~\eta~\cong~0.13 ~,
\end{equation}
with $T$ and $\mu_B$ fixed by hadron yield data.
The model requires rather large enhancement for the open charm
production (up to a factor of about $3.5$ within the rapidity window of
the NA50 spectrometer).
A direct measurement of the open charm in Pb+Pb at CERN SPS by NA60
would allow to test this prediction.

The SCM does not describe the NA50 data
at small values of
$E_T$ (the peripheral collisions with $E < 27$~GeV).
This can be also seen from the $\psi'$ data.
For $T\cong 170$~MeV the value of the thermal
ratio is
$\langle \psi^{\prime} \rangle /\langle J/\psi \rangle
\cong 0.04$.
This is in agreement with data in Pb+Pb collisions at SPS for
$N_p>100$, but is in contradiction with data in p+p, p+A and very
peripheral A+A collisions.
% \cite{psi'}.
This fact was first noticed in Ref.~\cite{psi'1}.

\section{$J/\psi$ enhancement in A+A Collisions at the RHIC}
The number of directly produced
$c\overline{c}$ pairs at the RHIC energies
can be estimated 
in the pQCD approach and 
used then as the input for the SCM.
The pQCD calculations for ${c\overline{c}}$ production
cross sections were
first done in Ref.\cite{comb}.  
For the cross section
$\sigma(pp\rightarrow
c\overline{c})$ of the charm production in p+p collisions we use 
the results presented in Ref.\cite{ruusk}.
This leads to the value of
$\sigma(pp\rightarrow
c\overline{c})~\cong 0.35$~mb at $\sqrt{s}=200$~GeV
and the $\sqrt{s}$-dependence of the cross section
for $\sqrt{s}=10\div200$~GeV is parameterized
as \cite{Go3}:
\begin{equation}\label{pert1}
\sigma(pp\rightarrow c\overline{c})~=~\sigma_0~\cdot 
\left(1- \frac{M_{0}}{\sqrt{s}}\right)^{\alpha}~
\left(\frac{\sqrt{s}}{M_{0}}\right)^{\beta}~,
\end{equation} 
with $\sigma_0 \cong 3.392 $~$\mu$b, $M_{0}\cong 2.984$~GeV, 
$\alpha \cong 8.185$ and $\beta \cong 1.132$.

The number of produced $c\overline{c}$
pairs in A+A collisions is proportional to
the number of primary N+N collisions, $N_{coll}^{AA}$,
which in turn is proportional to $N_p^{4/3}$ \cite{eskola}:
\begin{equation}\label{pert}
\langle N_{c\overline{c}} \rangle~ = ~N_{coll}^{AA}(N_p)~
\frac{\sigma(pp\rightarrow c\overline{c})}{\sigma_{NN}^{inel}}~
\cong ~ C~ \sigma(pp\rightarrow c\overline{c})~N_p^{4/3}~,
\end{equation}
where $\sigma_{NN}^{inel}\cong 30$~mb is the inelastic N+N cross sections,
$C\cong 11$~barn$^{-1}$.

The results of the SCM can be studied analytically 
according to Eqs.(\ref{Npsi1},\ref{Npsi2})
in the limiting cases of small and large numbers of 
$\langle N_{c\overline{c}}\rangle$.
For
$N_{c\overline{c}}<<1$
one finds:
\begin{equation}\label{lim1}
R~
\equiv~\frac{\langle J/\psi \rangle}
{\langle N_{c\overline{c}}\rangle}~
\cong~  \frac{4 N_{J/\psi}^{tot}}{N_O^2}~\sim~\frac{1}{V}~\sim~
\frac{1}{\langle \pi \rangle} ~\sim~
\left(\sqrt{s}\right)^{-1/2}~N_p^{-1}~,
\end{equation}
where we use the energy dependence of the
pion multiplicity per nucleon participant 
$\langle \pi \rangle/N_p \propto  (\sqrt{s})^{1/2}$ \cite{Ga}
which approximately works in the SPS--RHIC energy region.
The behavior (\ref{lim1}) corresponds to 
the $J/\psi$ suppression: the ratio $R$ 
decreases with increasing of both $\sqrt{s}$ and $N_p$.
This takes place at the SPS: this energy is still 
too ``low''
as $\langle N_{c\overline{c}}\rangle <1$ even in the most central Pb+Pb
collisions. However, the behavior of the $J/\psi$ to $N_{c\overline{c}}$
ratio is changed dramatically 
at the RHIC energies \cite{Go3} (see Fig.~2 and Ref.~\cite{Go3} for
details). 
\begin{center}
 \mbox{
  \parbox{6.0cm}{
   \includegraphics[width=6cm]{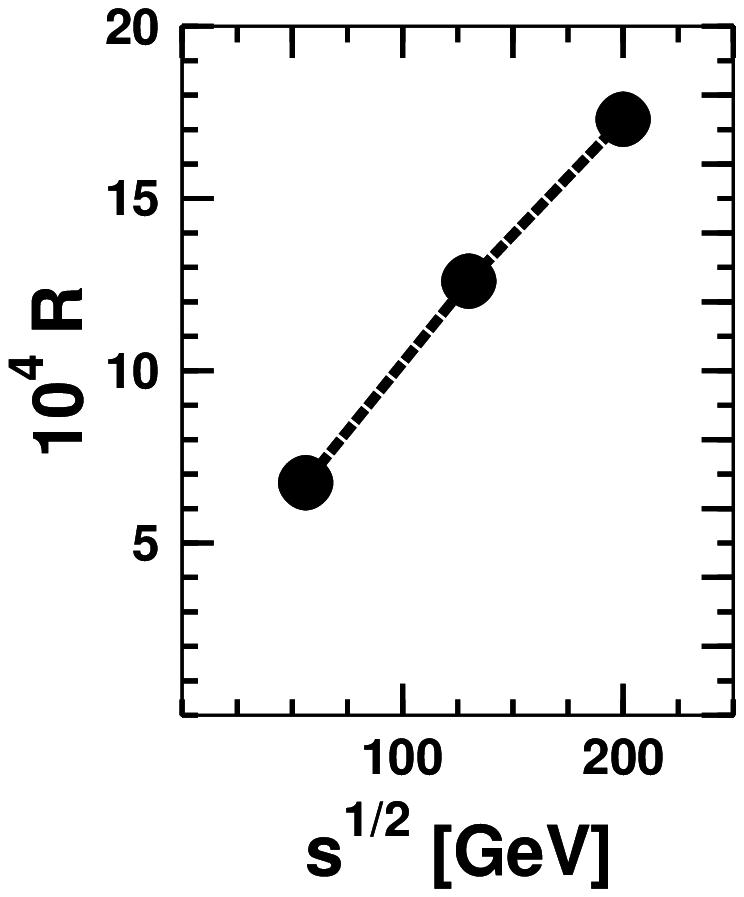}}
  \parbox{6.0cm}{
   \vspace*{-0.0cm}\includegraphics[width=6cm]{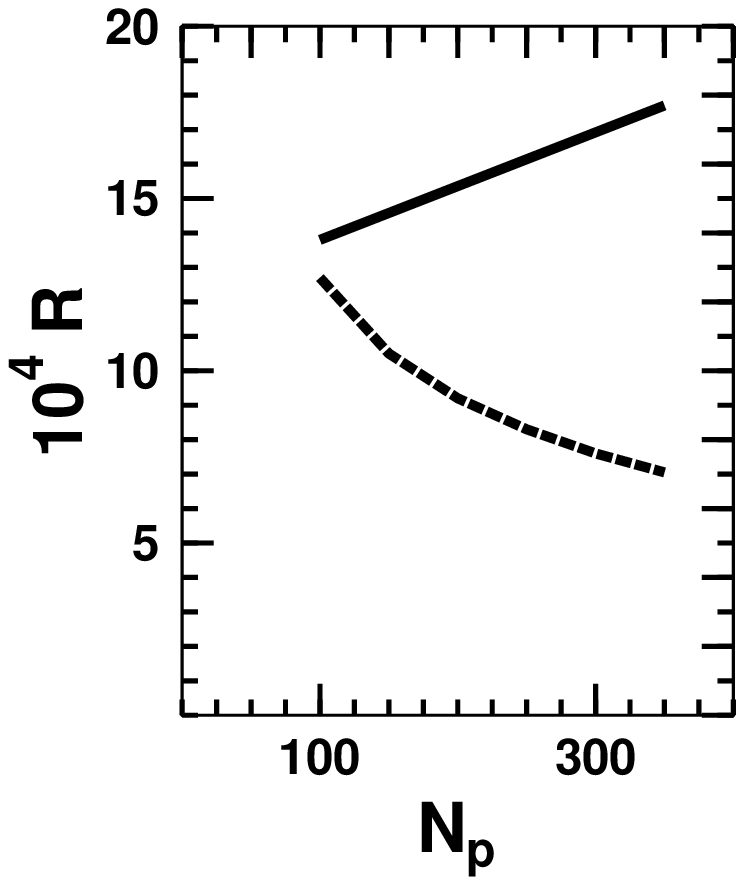}}
}
   \parbox{14cm}
{\footnotesize
Fig. 2. Left: 
The energy dependence of the $\langle J/\psi \rangle$ to
$\langle N_{c\overline{c}}\rangle$
ratio in central Au+Au collisions.
Points are the predictions of the SCM
for the RHIC energies: $\sqrt{s}=56, 130, 200$~GeV.
Right: 
 The $N_p$-dependence of the $\langle J/\psi \rangle$ to
$\langle N_{c\overline{c}}\rangle $
ratio. The lines are the predictions of the SCM. The dashed line
corresponds
to  $\sqrt{s}=56$~GeV (the $J/\psi$ {\it suppression}),
the solid line corresponds
to  $\sqrt{s}=200$~GeV (the $J/\psi$ {\it enhancement}).
}
\end{center}
In central Au+Au collisions 
at $\sqrt{s}=200$~GeV the expected value of $\langle N_{c\overline{c}}
\rangle$
is essentially larger than unit.  For $\langle
N_{c\overline{c}}\rangle >>1$
one finds ($\beta\cong 1.1$):  
\begin{equation}\label{lim2}
R~
\equiv~\frac{\langle J/\psi \rangle}{\langle N_{c\overline{c}}\rangle }~
\cong~  \frac{2
\langle
N_{c\overline{c}}\rangle}{N_O}~\sim~\frac{\langle 
N_{c\overline{c}}\rangle }{V}~\sim~
\frac{\langle N_{c\overline{c}}\rangle }{\langle \pi \rangle} ~\sim~
(\sqrt{s})^{\beta 
-1/2}~N_p^{1/3}~.
\end{equation}

%Eqs.~(\ref{lim1},\ref{lim2}) reveal a remarkable feature
%of the SCM: the $J/\psi$ to $\langle N_{c\overline{c}}\rangle $
%ratio reveal both the $J/\psi$ {\it suppression} (\ref{lim1})) 
%and $J/\psi$ {\it enhancement} (\ref{lim2}) behaviors.
%The measurements in Au+Au collisions at RHIC give a unique possibility 
%to check these predictions.
%The results of the SCM for the $J/\psi$ 
%to $\langle N_{c\overline{c}}\rangle $ ratio  are presented in
%Figs.~2 and 3.
%Both the suppression (the dashed line in Fig.~3) and enhancement
%(the solid lines in Figs.~2 and 3) behaviors are clearly seen.

\section{The $m_T$-spectra of $J/\psi$ and $\psi^{\prime}$ 
and QGP hadronization}
In Ref.\cite{BGG} we formulated the hypothesis  
that the kinetic freeze-out of $J/\psi$  and $\psi^{\prime}$ mesons takes
place directly at hadronization.
The effect of rescattering in the hadronic phase was recently studied  
within
a ``hydro + cascade'' approach \cite{BD,Sh}.  
A+A collisions are considered there to proceed in three stages:
hydrodynamic QGP
expansion (``hydro''),
transition from QGP to HG (``switching'')
and the stage of hadronic rescatterings and resonance decays
(``cascade'').
The switching from hydro to cascade takes place at $T=T_c$, where the
spectrum of hadrons leaving the surface of the QGP--HG transition is taken
as an input for the subsequent cascade calculations.
The results of Refs.~\cite{BD,Sh} suggest that
the transverse momentum spectra of $\Omega$ is only weakly affected during
the cascade stage \footnote{The corresponding calculation for charmonia
are not yet performed within this model.} and give therefore a
straightforward
measure of the collective hydro velocity at the `switching' surface
$T=T_c$.
In Ref.~\cite{GBG} we
demonstrated that   
in Pb+Pb collisions at 158~A$\cdot$GeV the $m_T$-spectra of $\Omega^{\pm}$
\cite{Omega} can be explained
simultaneously with the $m_T$-spectra of $J/\psi$  and $\psi^{\prime}$
mesons
\cite{mt}
using the same set
of the hadronization parameters.
%$T\cong 170$~MeV, $\overline{v}_T\cong
%0.2$.

Assuming the fluid freeze-out at constant temperature $T$,
the transverse mass spectrum of
$i$-th hadron species in cylindrically symmetric
and longitudinally boost invariant fluid expansion
equals approximately to (see Ref.~\cite{GBG} for further references and
details):
\begin{equation}\label{hydro1}
\frac{dN_i}{m_T dm_T}~
%\sim~m_T~
% K_1\left(\frac{m_T (1+\frac{1}{2} \overline{v}_T^2)}{T}\right)~
%I_0\left(\frac{p_T\overline{v}_T}{T}\right)~  
\propto ~
%\left(m_T\right)^{1/2}~
\sqrt{m_T}~
\exp\left(-~\frac{m_T (1+\frac{1}{2}\overline{v}_T^2)}{T}\right)
~I_0\left(\frac{p_T\overline{v}_T}{T}\right)~,
\end{equation}
where $m_T=(m_i^2+p_T^2)^{1/2}$ and $\overline{v}_T$ is the average
transverse flow velocity.
\begin{center}
   \includegraphics[width=10cm,height=9cm,angle=0]{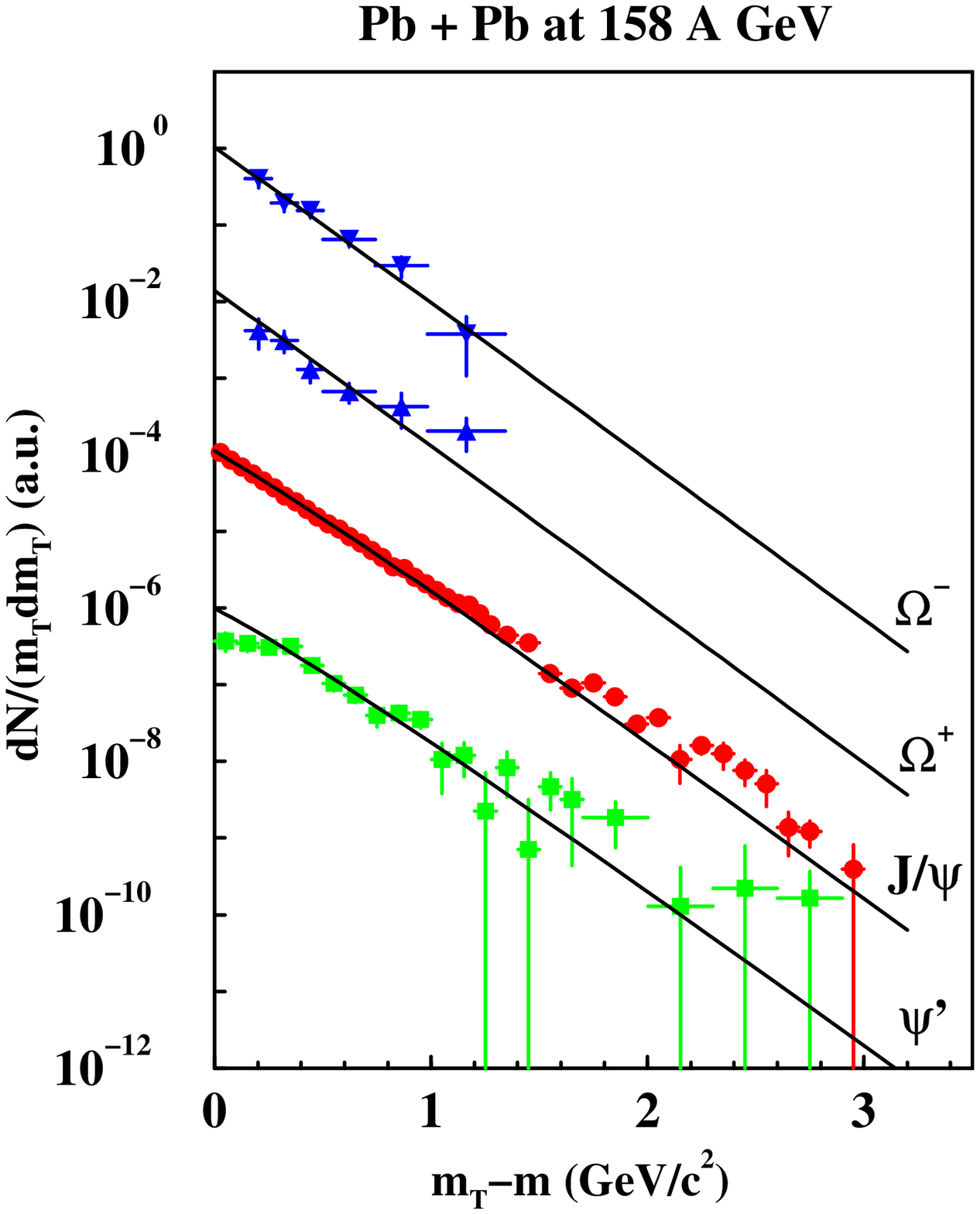}\\
   \parbox{14cm}
{\footnotesize
        Fig.~3:
The $m_T$-spectra
measured
at midrapidity in Pb+Pb  at 158~A$\cdot$GeV
 by WA97 \cite{Omega} for $\Omega$ 
and by NA50 \cite{mt} for $J/\psi$ and $\psi^{\prime}$
are presented (in arbitrary units) versus $m_T-m$.
The solid lines correspond to Eq.(\ref{hydro1}) with
$T=170$~MeV and $\overline{v}_T=0.19$.
} 
%}
\end{center}
%
%
%\vspace{0.3cm}
In Fig.~3 we present the fit with Eq.(\ref{hydro1}) of the measured
$m_T$-spectra.
The temperature is fixed as $T=170$~MeV and  $\overline{v}_T$
is considered as a free parameter. The shapes of all spectra are
simultaneously reproduced
at $\overline{v}_T= 0.19\pm 0.02$.

\section{ Conclusions}
%\begin{itemize}
%\item 
Statistical hadronization of the
QGP is probably an important
source of $J/\psi$ production \cite{Ga1}.
Within the SCM \cite{Br1,Go} the NA50 data   
on the $J/\psi$ production in Pb+Pb at 158~A$\cdot$GeV can be fitted
\cite{Go2}
(see Fig.~1)
for central collisions $N_p>100$.
A large enhancement of the open
charm over an extrapolation from the p+p data
is however required \cite{Go2,Go1}.
A direct measurement of the open charm in Pb+Pb at CERN SPS by NA60
would allow to test the SCM selfconsistency.
%\item 

The SCM predicts that the $J/\psi$ {\it suppression}
at the SPS 
%and in peripheral Au+Au collisions at lower RHIC energy
should be changed into the $J/\psi$
{\it enhancement} in central 
Au+Au collisions at the RHIC energies \cite{Go3} (see Fig.~2).
%\item

The shapes of $m_T$-spectra for $\Omega$, $J/\psi$ and $\psi^{\prime}$ are
simultaneously reproduced \cite{GBG} (see Fig.~3)
in the hydrodynamical picture of the QGP hadronization with 
$T=170$~MeV, $\overline{v}_T=0.19\pm 0.02$.
This supports the hypothesis that
formation and the kinetic freeze--out of charmonia occurs
at the hadronization \cite{BGG}.

%\end{itemize}

\vspace{0.3cm}
{\bf  Acknowledgments.}  
{\small I am thankful to K.A.~Bugaev, M. Ga\'zdzicki,
W.~Greiner, A.P.~Kostyuk, L.~McLerran and H.~St\"ocker for fruitful
collaboration. I am also 
thankful to F.~Becattini, L.~Bravina, P.~Braun-Munzinger, J.~Cleymans, 
 A.~Dumitru, L.~Gerland, D.~Kharzeev, I.N.~Mishustin, G.C.~Nayak,  
K.~Redlich, Yu.M.~Sinyukov, J.~Stachel, D.~Teaney and Nu Xu for
comments and discussions.
The financial support from the Humboldt Foundation is acknowledged.
The research described in this publication was made possible in part by 
Award \# UP1-2119 of the U.S. Civilian Research and Development
Foundation for the Independent States of the Former Soviet Union
(CRDF) and INTAS grant 00-00366}.


\begin{thebibliography}{99}
\itemsep=0cm
\bibitem{HG}   
J. Cleymans and H. Satz, Z. Phys. {\bf C57} (1993) 135;
J. Sollfrank, M.~Ga\'zdzicki, U. Heinz and J. Rafelski, Z. Phys. {\bf C61}
(1994) 659;
P.~Braun-Munzinger, I.~Heppe and J.~Stachel,
%``Chemical equilibration in Pb + Pb collisions at the SPS,''
Phys. Lett.  {\bf B465} (1999) 15;
%[nucl-th/9903010];\\
G.~D.~Yen and M.~I.~Gorenstein,
%``The analysis of particle multiplicities in Pb + Pb collisions at CERN
% SPS within hadron gas models,''
Phys. Rev. {\bf C59} (1999) 2788;
%[nucl-th/9808012].
F.~Becattini {\it et al}.,   
% J.~Cleymans, A.~Keranen, E.~Suhonen and K.~Redlich,
%``Features of particle multiplicities and strangeness production in
% central heavy ion collisions between 1.7-A-GeV/c and 158-A-GeV/c,''
%hep-ph/0002267.
 Phys. Rev. {\bf C64} (2001) 024901.
\bibitem{HG1}
N. Xu and M. Kaneta,  Nucl. Phys. {\bf A698},
(2002) 306;
P.~Braun-Munzinger {\it et al}.,
%D. Magestro, K. Redlich and J.~Stachel,
%hep-ph/0105229.
Phys. Lett. {\bf B518} (2001) 41;   
W. Frolkowski, W. Broniowski and M. Michalec, Acta Phys. Pol. {\bf B33}
(2002) 761.
\bibitem{Karsch}
%M. Oevers {\it et al}.,
%F. Karsch, E. Laermann and P. Schmidt,
%{\it Nucl. Phys. Proc. Suppl.} {\bf 73} (1999) 465;\\
F. Karsch,
Nucl. Phys. Proc. Suppl. {\bf 83-84} (2000) 14.
\bibitem{Satz1}
T.  Matsui and  H. Satz, {\it Phys. Lett.}  {\bf B178} (1986)  416.
\bibitem{Ga1}  
M. Ga\'zdzicki and M.I. Gorenstein,  Phys. Rev. Lett.
{\bf 83} (1999) 4009.
\bibitem{Br1}
 P. Braun-Munzinger and J. Stachel, Phys. Lett.  {\bf B490}
 (2000) 196.
\bibitem{Go}
M.I. Gorenstein,
% {\it et al}.,
A.P. Kostyuk, H. St\"ocker and  W. Greiner,
 Phys. Lett.  {\bf B498} (2001) 277,  J. Phys {\bf G27}
(2001) L47.
\bibitem{Go2}
A.P. Kostyuk,
% {\it et al}., 
M.I. Gorenstein, H. St\"ocker and  W. Greiner,
 hep-ph/0110269.
\bibitem{Go1}
A.P. Kostyuk, M.I. Gorenstein and W. Greiner,
Phys. Lett. {\bf B519} (2001) 207.
\bibitem{Go3}
M.I. Gorenstein {\it et al}., 
hep-ph/0012292,
Phys. Lett. {\bf B524} (2002) 265.
\bibitem{Raf1}
J. Rafelski, Phys. Lett.  {\bf B62} (1991) 333.
\bibitem{ce}
J. Rafelski  and  M. Danos, Phys. Lett.  {\bf B97}  279
(1980);
 K.~Redlich  and L.~Turko, Z. Phys.   {\bf C5} (1980)  541;
J.~Cleymans,  K.~Redlich  and  E.~Suhonen, 
 Z. Phys. {\bf C51} (1991) 137;
M.I.~Gorenstein,  M.~Ga\'zdzicki and  W.~Greiner,
 Phys. Lett. {\bf B483} (2000)  60.
%\bibitem{pdg}
%Particle Data Group,
%``Review of particle physics,''
%Eur. Phys. J.   {\bf C15} (2000) 1.
\bibitem{NA50}
M.C.~Abreu {\it et al.,}  (NA50),
Phys. Lett. {\bf B410} (1997) 327, {\bf B410} (1997) 337,
 {\bf B477} (2000) 28 
{\bf B450} (1999) 456.
\bibitem{Ko}
A.P. Kostyuk {\it et al}.,
% M.I. Gorenstein, H. St\"ocker and W. Greiner,
(in preparation).
%\bibitem{psi'}
%M. Gonin {\it et al.,   Presented at 3rd International
%Conference on Physics and Astrophysics of QGP} 1997 PRINT-97-208
\bibitem{psi'1}
H. Sorge, E. Shuryak and I. Zahed, Phys. Rev. Lett. {\bf 79} 
(1997) 2775.
\bibitem{comb}
B.L. Combridge, Nucl. Phys.  {\bf B151} (1979) 429.
\bibitem{ruusk}
P.L. McGaghey {\it et al.},
% E. Quack, P.V. Ruuskanen, R.~Vogt
%and X.-N. Wang, 
Int. J. Mod. Phys. {\bf A10} (1995) 2999.
\bibitem{Ga}
M. Ga\'zdzicki, private communication.
%
\bibitem{eskola}
K.J. Eskola, K. Kajantie  and J. Lindfors 
Nucl. Phys.  {\bf B323} (1989)  37.
%
\bibitem{BGG}
K. A. Bugaev, M. Ga\'zdzicki and M. I. Gorenstein,
Phys. Lett. {\bf B523} (2001) 255;
K. A. Bugaev, nucl-th/0112016.
%
\bibitem{BD}
S. Bass and A. Dumitru, Phys. Rev. {\bf C61} (2000) 064909.
% 
\bibitem{Sh}
D. Teaney, J. Lauret and E.V. Shuryak, nucl-th/0110037.
% 
\bibitem{GBG}
M.I. Gorenstein, K.A. Bugaev and M. Ga\'zdzicki,
hep-ph/0112197.
%
\bibitem{Omega}
F. Antinori {\it et al}., (WA97), J. Phys. {\bf G27} (2001) 375.
%
\bibitem{mt}
M.C. Abreu {\it et al}., (NA50), Phys. Lett. {\bf B499} (2001) 85.
%
%\bibitem{Heinz}
%E. Schnedermann, J. Sollfrank and U. Heinz,
%Phys. Rev. {\bf C48} (1993) 2462.
%
%\bibitem{Omega1}
%E. Anderson et al., (WA97),  Phys. Lett. {\bf B433} (1998) 209.

\end{thebibliography}
\end{document}